\documentclass[aps,prl,twocolumn,citeautoscript]{revtex4-1}

\bibpunct{[}{]}{,}{n}{}{}

\usepackage{verbatim}
\usepackage{hyperref}

\usepackage{multirow}
\usepackage{graphicx}
\usepackage{longtable}
\usepackage[utf8]{inputenc}
\usepackage{epstopdf}
\usepackage{longtable}
\usepackage{textcomp}
\usepackage[usenames,dvipsnames]{color}
\usepackage{amsbsy}
\usepackage{amsmath}
\usepackage{amsfonts}
\usepackage{mathtools}
\usepackage{braket}
\usepackage{array}
\usepackage{gensymb}
\usepackage[normalem]{ulem}
\usepackage{placeins}

\newcommand{\eg}{\textit{e.g.},\@ }

\newcommand{\ie}{\textit{i.e.},\@ }

\newcommand{\vc}[1]{\boldsymbol{\mathrm #1}}

\newcommand{\abs}[1]{\lvert#1\rvert}
\newcommand{\matt}[1]{\mathcal{#1}}

\makeatletter
  \def\my@tag@font{\normalsize}
  \def\maketag@@@#1{\hbox{\m@th\normalfont\my@tag@font#1}}
  \let\amsmath@eqref\eqref
  \renewcommand\eqref[1]{{\let\my@tag@font\relax\amsmath@eqref{#1}}}
\makeatother

\allowdisplaybreaks

\begin{document}

\title{Atomistic perspective of long lifetimes of small skyrmions at room temperature}

\newcommand{\fz}{Peter Gr\"unberg Institut and Institute for Advanced Simulation, Forschungszentrum J\"ulich and JARA, 52425 J\"ulich Germany}
\author{Markus Hoffmann}
\email{m.hoffmann@fz-juelich.de}  
\author{Gideon P. M\"uller}
\affiliation{\fz}
\author{Stefan Bl\"ugel}
\affiliation{\fz}

\date{\today}

\begin{abstract}
  
 The current development to employ magnetic skyrmions in novel spintronic device designs has led to a demand for room temperature-stable skyrmions of ever smaller size.
 We present extensive studies on skyrmion stability in atomistic magnetic systems in two- and three-dimensional geometries.
 We show that for materials described by the same micromagnetic parameters, the variation of the atomistic exchange between different neighbors, 
 the stacking order, and the number of layers of the atomic lattice can significantly influence the rate of the thermally activated decay of a skyrmion.
 These factors alone are important considerations, but it is shown that their combination can open up novel avenues of materials design in the search for sub-$10$~nm skyrmions, as their lifetime can be extended by several orders of magnitude.
\end{abstract}

\maketitle

Magnetic skyrmions~\cite{bogdanov1989thermodynamically,Muhlbauer915} are localized chiral magnetic textures on the nanometer scale.
They have become a timely and vital research topic due to their promising properties for future data storage and processing purposes~\cite{fert2013skyrmions}.
The expectations include the creation of devices with high storage densities and low operating energy costs.
To be viable as a competitive technology, these skyrmionic devices require stable sub-$10$~nm skyrmions at room temperature~\cite{sub10nm}.
This moves the importance of the thermal stability of single skyrmions into the focus of research.
The search for suitable material compositions~\cite{Wilde1701704, PhysRevB.100.014425, oike2016Interplay}, which fulfill the requirements of long skyrmion lifetimes, plays a major role in material system design.
A relation of materials' properties to lifetime does not currently exist.

One of the most commonly taken approaches relating materials' properties to magnetic ones  is the micromagnetic model~\cite{brown1963micromagnetics},
\begin{equation}
    \mathcal{H} = \int \text{d}\vc{r} \ \left[
        \left(\nabla\vc{m}\right)\matt{A}\left(\nabla \vc{m}\right)
        \ +\ \matt{D}:\matt{L}(\vc{m}) - \vc{m}\,\matt{K}\,\vc{m} \right]\,,
\label{eq:micro}
\end{equation}
which allows for both powerful numerical and elegant analytical calculations.
The first term of the micromagnetic Hamiltonian represents the exchange contribution via the spin stiffness $\matt{A}$, in its most general form a tensorial quantity, the second term is the skyrmion stabilizing Dzyaloshinskii-Moriya interaction (DMI) with the spiralization tensor $\matt{D}$ and the Lifshitz tensor $\matt{L}(\vc{m})=\nabla \vc{m}\times\vc{m}$~\cite{hoffmann2017}, $\matt{K}$ is a uniaxial anisotropy contribution, and $\vc{m}(\vc{r})$ is a continuous, normalized magnetization field.
Note that for most systems, symmetry arguments reduce the degrees of freedom in $\matt{A}$,  $\matt{D}$, and  $\matt{K}$~\cite{bogdanov1989, hoffmann2017}.
Additional terms that are often included are the magnetic field in form of a Zeeman term and the demagnetization field.
According to the micromagnetic model, sub-$10$~nm skyrmions at room temperature can only be stabilized by the DMI and in materials with low saturation magnetization, \eg in synthetic antiferromagnets or ferrimagnets~\cite{buttner2018theory,SciPostPhys.4.5.027}. Thus,
in case of small skyrmions in a ferromagnetic background and in the limit of thin films, the
 latter is often neglected in numerical simulations, since the leading order contribution to energetics can be captured by renormalizing the anisotropy constants~\cite{bogdanov1994properties}.
In the present study we therefore concentrate on the interactions given in Eq.~\eqref{eq:micro}.

With the reduction of the skyrmion size and the investigation of the skyrmion lifetime, the details of the underlying lattice and the details of the atomistic interactions begin to play an important role. The skyrmion annihilation goes along with a change of the topological charge accompanied by a reduction of the skyrmion size and a discontinuous magnetization transition across a saddle point, which is reflected by large-angle changes of the magnetization between neighboring atoms.    
In this context, it is important to remember that most skyrmions are studied in metallic magnetic systems, \eg in magnetic bulk  transition metals, in films or heterostructures, which additionally contain heavy metal interlayers or substrates. These systems have complex Fermi surfaces~\cite{Zimmermann:16}, which can lead to long-ranged oscillatory and anisotropic atomistic Rudermann-Kittel-Kasuya-Yosida (RKKY)-type exchange, $J_{ij}$,  or DM interactions, $\vc{D}_{ij}$, where the size and the sign of the magnetic interactions between pairs of magnetic moments on atom sites $i, j$ change with distance and direction ending up with competing interactions on the atomic scale.
Such competing interactions can substantially reduce the energy cost of large-angle spin rotations at the transition state.

Compared to the detailed atomistic information, the micromagnetic model is subject to a massive reduction of information. The micromagnetic quantities in \eqref{eq:micro} can be understood as a contraction of the atomistic information over the lattice points according to the relations~\cite{Schweflinghaus:16}
\begin{equation}
    \matt{A} = \frac{1}{2V}\sum_{\braket{i,j}} J_{ij} \vc{R}_{ij}\otimes \vc{R}_{ij}   
    \text{  and  }
    \matt{D} = - \frac{1}{V}\sum_{\braket{i,j}} \vc{D}_{ij}\otimes \vc{R}_{ij}\, ,
    \label{eq:relation}
\end{equation}
where the $\vc{R}_{ij}$ are the distance vectors connecting interacting sites.
With this reduction of information the micromagnetic model provides an excellent basis for systematic analyses as well as analytical investigations, but materials with potentially very different behaviors at skyrmion saddle points may all be described by the same micromagnetic parameters. 
Thus, important physical aspects such as the role of competing exchange interactions on large angle-rotation or skyrmions lifetime are insufficiently addressed by an effective stiffness and spiralization.

In this letter we go beyond the micromagnetic approach and determine the lifetime of skyrmions on the basis of an atomistic spin model.
We demonstrate a tremendous dependence of the rate of skyrmion decay on the details of nearest and next-nearest neighbor exchange and DM interactions by varying the atomistic parameters such that the micromagnetic parameters remain unchanged.
We then consider the commonly investigated and technologically important (111) textured multilayer systems and show how the decay depends on the number of monatomic hexagonal magnetic layers and the stacking order.
By combining these effects, we show how the lifetime of skyrmions can be changed by several orders of magnitude, offering a huge space of opportunities for materials design of skyrmions with large lifetime that cannot be covered by the micromagnetic theory and has been overlooked in the past.

\medskip\paragraph{Atomistic Spin Model}
We employ a spin-lattice model with classical vector spins $\vc{S}_i$ of unit length placed on atomic sites $i, j$  consisting of the Heisenberg model extended by the DMI  and uniaxial magnetic anisotropy energy
\begin{align}
    \mathcal{H} =- \sum_{\braket{i,j}}\! J_{ij}\, \vc{S}_i \cdot \vc{S}_j -  \sum_{\braket{i,j}}\vc{D}_{ij} \left(\vc{S}_i \times \vc{S}_j\right) + \sum_i K \left(S_i^z\right)^2\hspace{-0.1cm}
\label{eq:atomistic}
\end{align}
with the corresponding atomistic pair $J_{ij}$, $\vc{D}_{ij}$, and on-site $K$ interactions.
$\braket{i,j}$ denotes the summation over unique interaction pairs.

\medskip\paragraph{Atomistic Calculations}
Since skyrmions are subject to topological protection, thermally activated 
transitions from the metastable skyrmion  to the ferromagnetic ground state are typically rare events on the time scale of oscillations of the magnetic moments. This scenario is described by the transition state theory~\cite{wigner1938transition} adopted to multiple spin degrees of freedom~\cite{bessarab_harmonic_2012}.
For a given transition from an initial local minimum, $\mathrm{M}$, across a first-order saddle point, $\mathrm{S}$, of the energy landscape located on a path connecting the initial- and final-state minima, employing the harmonic approximation to the transition-state theory (HTST), the rate constant $\Gamma$ is given by an Arrhenius expression 
\begin{equation}
    \Gamma^\mathrm{HTST}(T) =
        \Gamma_0(T) \,
        e^{-\Delta E/k_\mathrm{B}T}\; ,
\label{eq: lifetime contributions}
\end{equation}
where $k_\mathrm{B}$ is the Boltzmann constant, $\Delta E=E^{\mathrm{S}}-E^{\mathrm{M}}$ is the activation energy, 
and $\Gamma_0(T)$ is a temperature-dependent preexponential factor. The latter depends on the ratio of all non-zero eigenmodes
determined through the harmonic approximation to the energy surface  at $\mathrm{M}$ and $\mathrm{S}$, respectively (see supplementary material for details~\cite{supplement}).
The deviation of the ratio of eigenvalues from unity can be interpreted as entropic narrowing of the spin degrees of freedom when the skyrmion passes through the saddle point region~\cite{desplat_thermal_2018, von_malottki_skyrmion_2018,desplat2019path}.
The temperature-dependence is given by $\Gamma_0 \propto \sqrt{T}^{\left(N_0^\mathrm{M} - N_0^\mathrm{S}\right)}$, where
$N_0^\mathrm{M}$ and $N_0^\mathrm{S}$ are the number of translationally invariant zero-energy spin excitation eigenmodes. The number of modes depends for example on the geometry and the skyrmion annihilation mechanism.
Thus, the two quantities determining the stability of a meta-stable skyrmion are the minimum activation energy, $\Delta E$, equivalent to the energy barrier for skyrmion annihilation  along a minimum energy path connecting the initial and final state (the inverse path relates to nucleation), and the prefactor $\Gamma_0$, which we will calculate and analyze in the following for the case of single isolated skyrmions.
For simplicity, we focus on one particular decay mechanism, namely the radial collapse, which is typically a dominant one for skyrmions in film geometry~\cite{bessarab_lifetime_2018, muller_duplication_2018}.

All atomistic calculations presented in this letter are based on the energy landscape described by the atomistic model \eqref{eq:atomistic} and were performed with \textit{Spirit}~\cite{spirit, muller_spirit_2019}.
This includes in particular the calculation of metastable and stable spin textures, their energies, and minimum energy paths using the Geodesic Nudged Elastic Band (GNEB) method~\cite{bessarab2015method} to obtain the saddle point configuration, as well as the calculation of eigenmodes and transition rates relying on the HTST~\cite{bessarab_harmonic_2012}.

In order to demonstrate the potential 
of atomistically resolved exchange interactions for the lifetime of skyrmions, we first focus on a ferromagnetic ($\mathcal{A}>0$, $J_1> 0$ and $J_2 > -1/3\, J_1$) monoatomic hexagonal lattice of nearest (NN) and next-nearest neighbor (NNN) interactions. We vary the exchange parameters $J_1$ and $J_2$ and the DM parameters, $\vc{D}_1$ and $\vc{D}_2$ such that the micromagnetic quantities, spin stiffness $\mathcal{A}$ and spiralization $\mathcal{D}$ in Eq.~\eqref{eq:relation}, remain constant.
Thus, all these systems are indistinguishable on the level of the micromagnetic model.
A brief study of the influence on an additional third-nearest neighbor interaction can be found in the supplementary material~\cite{supplement}.

With the aim of describing skyrmions in prototypical ultrathin-films, we chose the micromagnetic parameters $\mathcal{A}=12.5\, {\mathrm{pJ}}/{\mathrm{m}}$ and  $\mathcal{D} = 10\, {\mathrm{mJ}}/{\mathrm{m}^2}$. 
Assuming a lattice parameter of $0.27$\,nm, as it is typical for Pt or Ir based systems, and a layer thickness of $0.2$\,nm, the chosen parameters translate to the  atomistic parameters $J_1=18$\,meV and $\abs{\vc{D}_1}=2$\,meV per pair of atoms for the case of purely NN interactions, $J_2=0$,  $\abs{\vc{D}_2}=0$, consistent with first-principles results based on density functional theory~\cite{dupe2014tailoring,meyer_isolated_2019} resulting in a  skyrmion diameter below $4$\,nm consistent with experiment~\cite{Romming:2015}.

\begin{figure}
    \centering
    \includegraphics[width=0.95\linewidth]{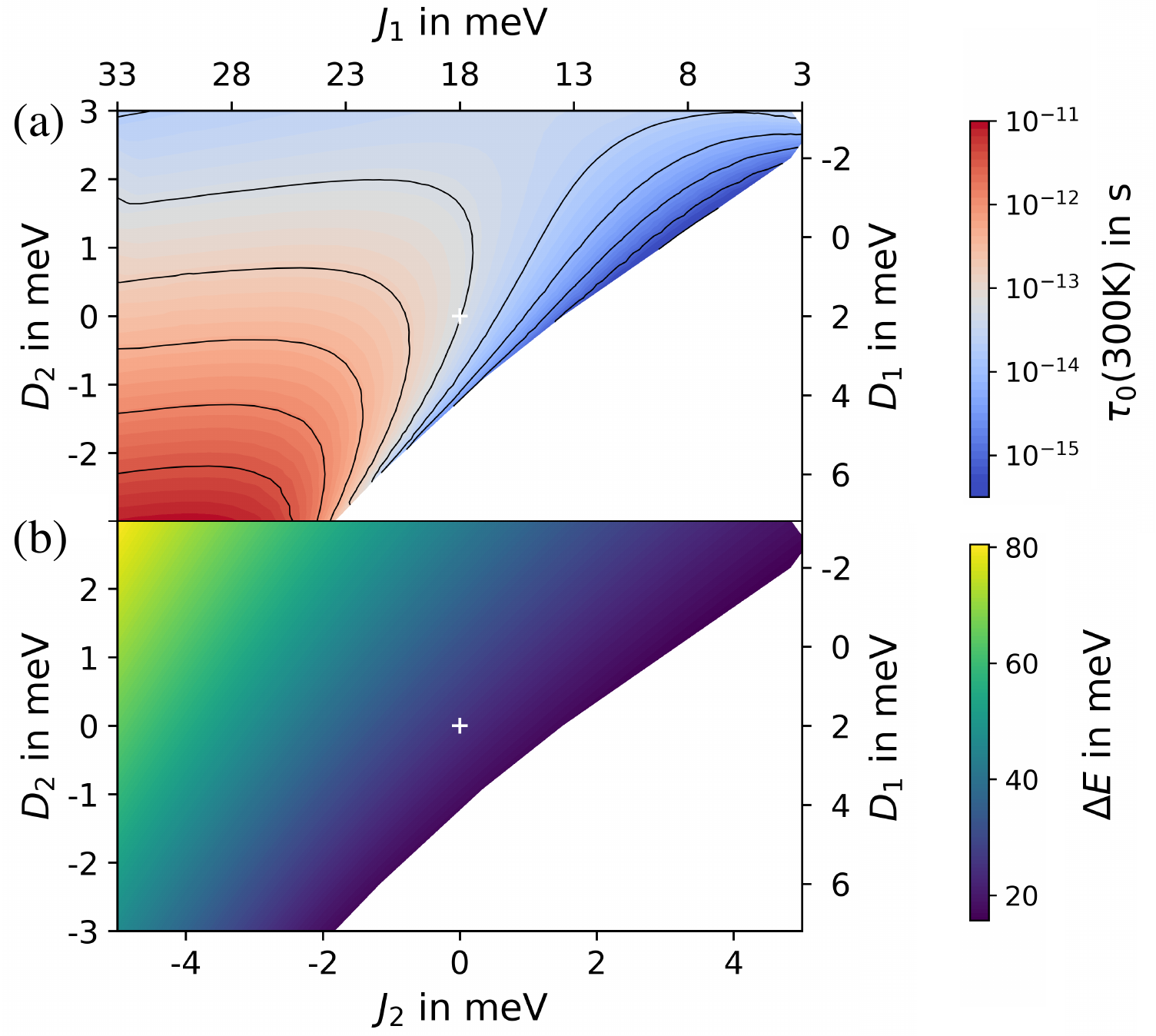}
    \caption{
    (a) Lifetime prefactor at room temperature, $\tau_0(300\mathrm{K}) = 1/\Gamma_0(300\mathrm{K})$, and (b) energy barrier $\Delta E$ for the radial collapse of a single skyrmion in a monolayer system in dependence on the NN and NNN contributions of both exchange (x-axis) and DM (y-axis) interactions. Note the linear dependence of the attempt rate on the temperature.
    The case of $J_2 = \abs{\vc{D}_2} = 0$, \ie pure NN interactions, is highlighted by a white cross.
    Note that the ratio of $J_2/J_1$ is given by $\mathcal{A}=12.5\, {\mathrm{pJ}}/{\mathrm{m}}$ and Eq.~\eqref{eq:relation}.
    }
    \label{fig:j2d2}
\end{figure}

Staying within the ferromagnetic regime, we vary the NNN interactions, $J_2$ from $-5$ to $5$\,meV and $\abs{\vc{D}_2}$ from $-3$ to $3$\,meV, and adjust the NN interactions while keeping $\mathcal{A}$ and $\mathcal{D}$ constant according to Eq.~\eqref{eq:relation}, \ie $J_1=18\,\text{meV}-3\,J_2$ and ${\abs{\vc{D}_1}=2\,\text{meV}-\sqrt{3}\, \abs{\vc{D}_2}}$.
Fig.~\ref{fig:j2d2} shows the calculated energy barrier $\Delta E$ and lifetime prefactor  $\tau_0 = 1/\Gamma_0$.
Two results are immediately evident:
(i) a significant dependence of both quantities on the interplay between NN and NNN interactions and
(ii) while the energy barrier shows a monotonous dependence, the lifetime prefactor depends in a more intricate way on the variation
of the atomistic parameters \footnote{Even though the micromagnetic parameters remain the same, atomistic simulations exhibit a small dependence of the radius of the  energy minimizing skyrmions  on the ratios of $J_2/J_1$ and $D_2/D_1$.
However, these variations are only in the order of 10\%, so that their entropic contributions to the skyrmion lifetime can be expected to be small.}.
While the change in the energy barrier is mainly a consequence of the energy change of the saddle point configuration by the change of the exchange interaction, the lifetime prefactor is influenced by changes in both the skyrmion state as well as at the saddle point.
Its complex dependence can be understood by significantly distinct behaviors of different eigenmodes at the skyrmion state M, such as the skyrmion's breathing and elliptical modes, with respect to the ratios of NN and NNN interactions (see supplementary material~\cite{supplement}). The plot has no entries in the $J$-$D$-region where the saddle point configuration changes.
A different saddle point configuration corresponds to  a different transition mechanism, which we exclude for the sake of conciseness.

\begin{figure}[t]
    \centering
    \includegraphics[width=\linewidth]{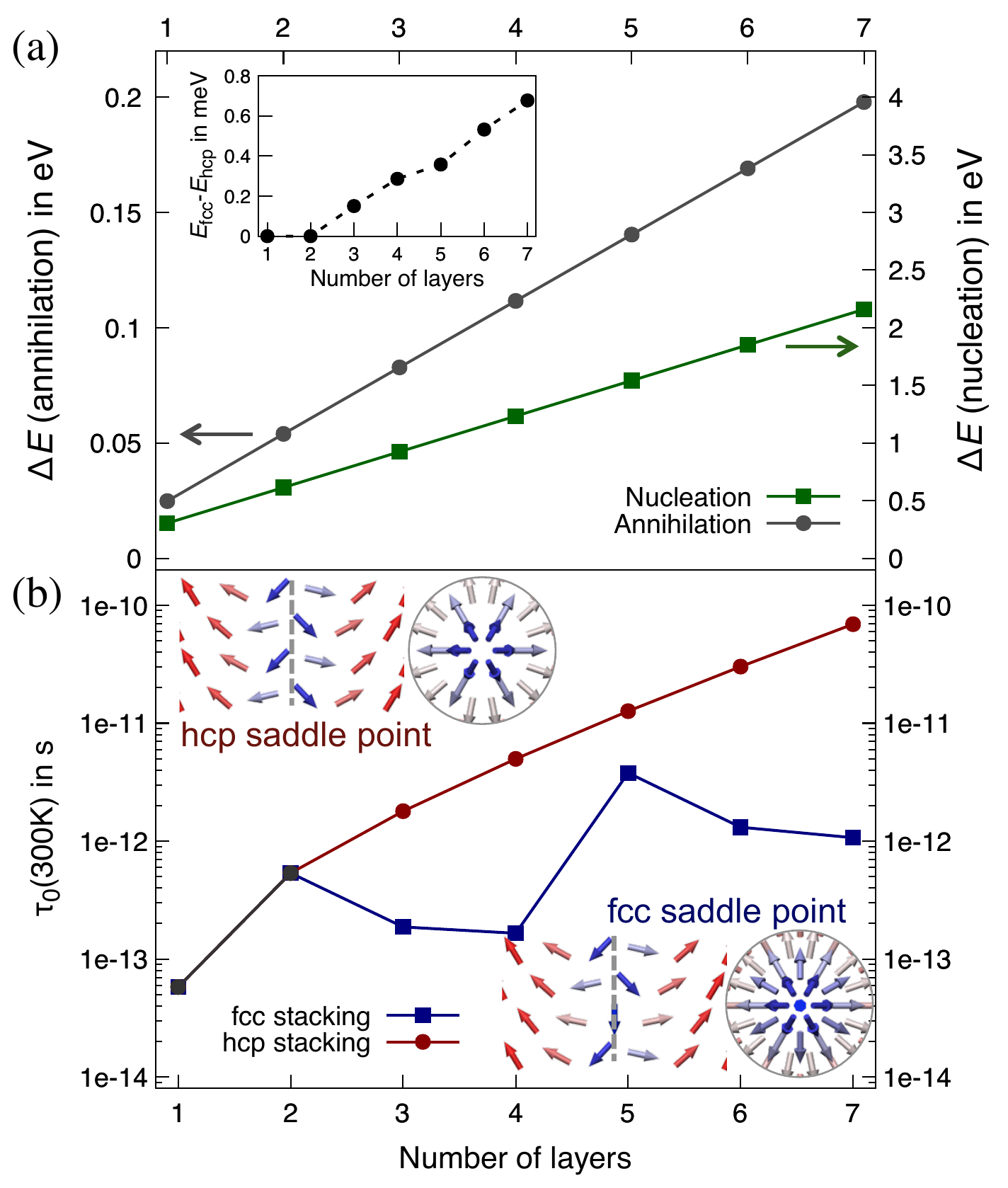}
    \caption{
    (a) Energy barrier for the annihilation (gray circles) and the nucleation (green squares) of a skyrmion in dependence on the number of magnetic layers.
    The energy difference between the saddle point energies of fcc and hcp stacked layers is shown in the inset.
    (b) Lifetime prefactor at room temperature, $\tau_0(300\mathrm{K}) = 1/\Gamma_0(300\mathrm{K})$, in dependence on the number of magnetic layers for hcp (red) and fcc (blue) stacking orders, calculated from HTST. Note the linear dependence of the attempt rate on the temperature.
    The insets show side- and top-views of the transition states for the four-layer system in fcc and hcp stacking, respectively.
    The center of the skyrmionic structure is highlighted by a dashed line. Parameters for both plots are $J_1=18$~meV, $D_1=2$~meV and $K=0.7$~meV.
    }
    \label{fig:combined}
\end{figure}

Now we turn to thicker ferromagnetic (111) textured films, for whose description we consider two limiting cases: fcc and hcp stacked layers of two-dimensional hexagonal films. Here, the atomistic calculations shine light onto the impact of the crystal lattice and film thickness on the  skyrmion lifetimes.
Fig.~\ref{fig:combined}(a) displays the dependence of the skyrmion nucleation and annihilation barriers on the sample thickness.
Both exhibit a linear dependence, whereby the transition mechanism corresponds to  a radial collapse homogeneous over all layers.
While this suggests increasing the sample thickness to gain skyrmion stability, the onset of a different annihilation mechanism limits the potential of this approach in thicker films.
Thermal excitations may nucleate a Bloch point or pair of Bloch points in the skyrmion, which can then propagate along the skyrmion tube to collapse it, resulting in an upper bound for the maximally achievable energy barrier~\cite{muller_spirit_2019}.
Especially for thick films this may happen at any point along the skyrmion, severely limiting its stability.

While the effect of stacking order on the energy barrier is less than 1~meV (see inset of Fig.~\ref{fig:combined}a) and thus negligible, the lifetime prefactor turns out to be strongly modified.
As Fig.~\ref{fig:combined}(b) shows, the lifetime prefactor of the fcc stacking is significantly decreased with respect to the hcp stacking for $3$ layers or more.
The step-like dependency and especially the periodicity of three layers can be understood by the fact that the skyrmion at the saddle point along the collapse path becomes significantly smaller in size and experiences the underlying lattice. Therefore, while the minimum exhibits two zero-energy translation modes, $N_0^\mathrm{M}=2$, at the saddle point the lattice is felt as corrugations in the energy landscape and no zero-energy modes remain, $N_0^\mathrm{S}=0$. This implies a temperature-dependence $\Gamma, \Gamma_0 \propto T$.
In fact, the energy is minimized when the skyrmion is positioned in a hollow-like position with respect to the underlying lattice as atoms with local moments pointing opposite to the ferromagnetic background magnetization is avoided.
This produces  a significant difference between the hcp and fcc stacking.
While the hcp stacking naturally permits such a position, in the fcc case there is no position without an atom at the center, as illustrated in Fig.~\ref{fig:combined}b.
An additional atom appears in the center of the configuration at every third layer, causing the observed periodicity in the lifetime prefactor, as well as the enhanced energy barrier compared to the hcp case (see Fig.~\ref{fig:combined}a).
The effect on the lifetime prefactor is significantly more pronounced due to the influence of the translational modes, for which we found the same step-like dependence in the eigenvalue.
This shows that the lattice symmetry can have a deep impact on the lifetime of metastable magnetic states.

\begin{figure}[t]
    \centering
    \includegraphics[width=0.9\linewidth]{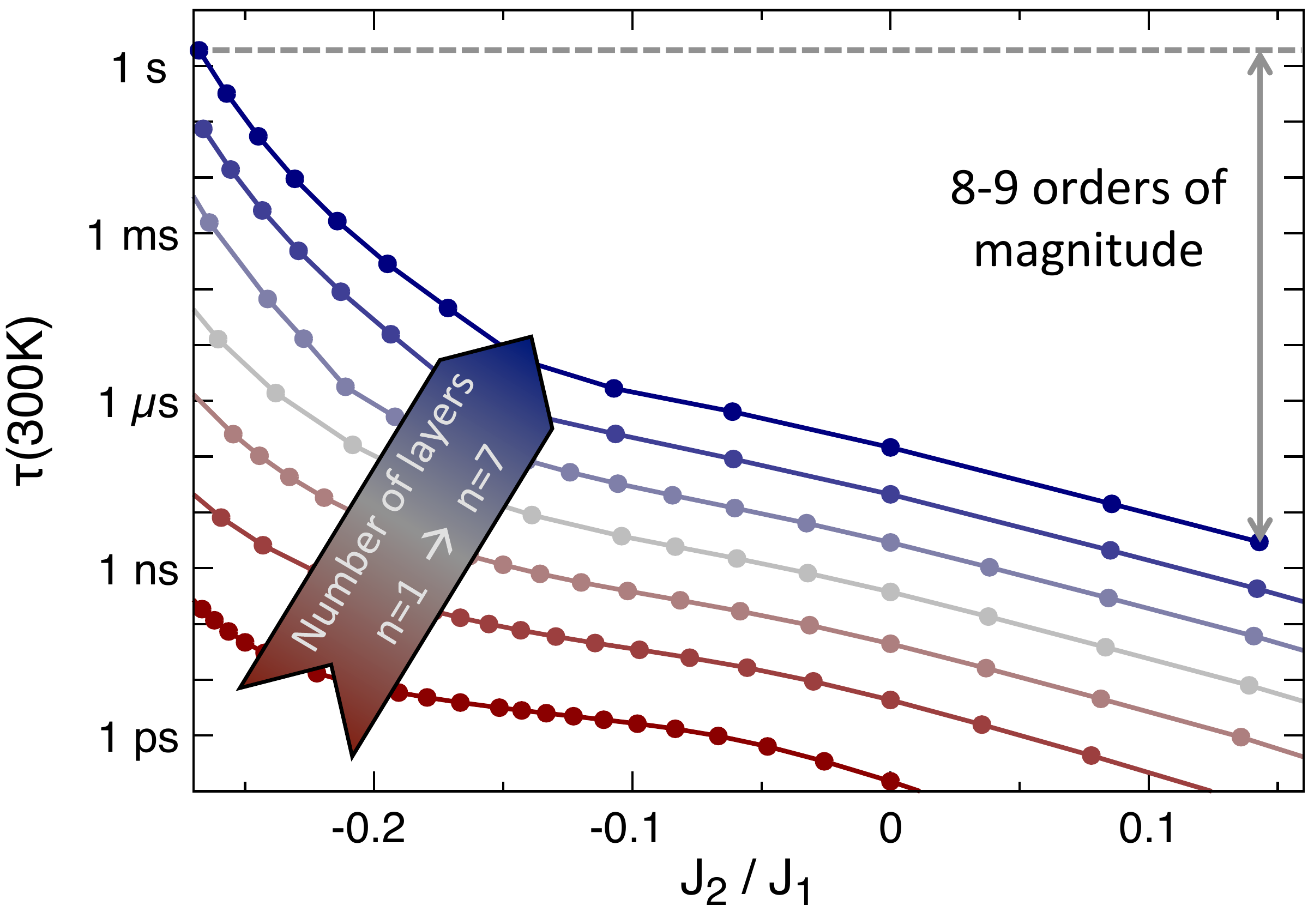}
    \caption{
        Estimated room temperature lifetime $\tau(300\mathrm{K}) = 1/\Gamma(300\mathrm{K})$ in dependence on the ratio of exchange interactions $J_2/J_1$, shown for $1$ (red) to $7$ (blue) atomic layers.
        Note that, as before, the absolute values of $J_1$ and $J_2$ are given by $\mathcal{A}=12.5\, {\mathrm{pJ}}/{\mathrm{m}}$ and Eq.~\eqref{eq:relation}.
        While the lifetime for one or a few layers already varies over several orders of magnitude over the shown interaction ratios, at seven layers this variation spreads over $9$ orders of magnitude.
    }
    \label{fig:lifetime}
\end{figure}

Considering the combination of the above findings we investigated the impact of systematically varying NN and NNN-exchange interactions on the lifetime of a single magnetic skyrmion at room temperature for the case of hcp stacking as a function of the number of atomic layers.
The results, shown in Fig.~\ref{fig:lifetime}, demonstrate extreme changes in the lifetime over the ratio of both parameters.
At $7$ layers, the shown range of interaction ratios can change the lifetime by $9$ orders of magnitude, while the case of $J_2/J_1 = -0.26$ \footnote{In three-dimensionally stacked hexagonal layers, $J_1$ connects NN atoms in the plane and neigboring planes, $J_2$ connects NNN between neigboring planes, and the NNN atoms in the plane are actually third-nearest neighbor atoms. To be consistent with the two-dimensional system in our analyses and to keep the number of parameters small, we set $J_2=0$ and continue to name the interactions between the NNN atoms in the plane as $J_2$.} shows the same range over thicknesses between $1$ and $7$ layers.

A subtlety arises in the treatment of the translational modes at elevated  temperatures~\cite{bessarab_lifetime_2018, supplement}.
When the thermal fluctuations are significantly larger than the energy barriers related to the translation of the saddle point configuration, $\lambda_\mathrm{tr}V_0 \ll k_\mathrm{B}T$, they effectively become zero modes~\cite{von_malottki_skyrmion_2018}.
This can have an impact on the lifetime prefactor $\Gamma_0$, shown in Fig.~\ref{fig:combined}b, as well as the temperature dependence of the lifetime, which depends on the numbers of zero modes at the minimum and saddle point (see Eq.~\eqref{eq: lifetime contributions}).
By comparing to calculations where all translational modes were treated as zero modes, we estimate an increase of the lifetime at room temperature by an order of magnitude.
Our results presented in Fig.~\ref{fig:lifetime} therefore represent a conservative estimate.

\medskip\paragraph{Conclusion}

We have shown that an atomistic description offers new perspectives on the lifetimes of skyrmions that cannot be obtained from the typically used standard micromagnetic model. For example, we could show that the competing interaction between nearest and next-nearest neighbors, the stacking order of hexagonal atom layers and the sample thickness  can change the lifetimes at room temperature by many orders of magnitude, although they are identical from the point of view of the micromagnetic description. Considering that in real metallic multilayer systems there are many competing interactions between atomic pairs of very different distances, there is a huge potential for lifetime design, \eg employing concepts from materials informatics.

Consistent with our results, previous studies comparing two particular parameter sets have found frustration to be able to enhance skyrmion stability~\cite{von_malottki_enhanced_2017,von_malottki_skyrmion_2018}.
Our study significantly expands upon this notion and opens a new avenue for engineering skyrmions in thin films.

While this study shows lifetimes of up to $1$\,s for a $4$\,nm skyrmion, we conjecture that $10$\,nm skyrmions may be stabilized on time scales relevant for technological applications since larger skyrmions tend to have higher energy barriers~\cite{varentsova_interplay_2019, potkina_antiskyrmions_2019} and can be more stable due to entropic effects~\cite{desplat_thermal_2018, von_malottki_skyrmion_2018,desplat2019path}. 

Our study focuses on thin films, but naturally bears consequences for more complicated magnetic systems.
In particular, the design of multilayer systems~\cite{jiang2017skyrmions} can benefit equivalently from our findings.
The intralayer behavior dominates the stability properties in such systems due to the weak coupling between the layers
and should be optimized accordingly.

\section*{Acknowledgments}
M.H. and S.B.\ acknowledge funding from
the DARPA TEE program through grant MIPR\# HR0011831554 from DOI. G.\,P.\,M.\ acknowledges funding from the Icelandic Research Fund (grant no.\ 185405-051).

\end{document}